\documentclass[useAMS,usenatbib]{mn2e}
\usepackage{graphicx}
\def\aap{A\&A}

\def\apj{ApJ}

\def\apjl{ApJ}

\def\mnras{MNRAS}

\def\aapr{A\&ARev}

\def\jrasc{JRASC}

\title[The long period eccentric orbit of HD\,167971]{The long period eccentric orbit of the particle accelerator HD\,167971 revealed by long baseline interferometry\thanks{Based on observations collected at the European Southern Observatory, Paranal, Chile under the program IDs: 381.D-0095, 086.D-0586 and 087.D-0264}}
\author[M. De Becker et al.]{M. De Becker$^{1}$\thanks{E-mail:
debecker@astro.ulg.ac.be}, H. Sana$^{2}$, O. Absil$^{1}$\thanks{Postdoctoral Researcher F.R.S.-FNRS (Belgium)}, J.-B. Le Bouquin$^{3}$ and R. Blomme$^{4}$\\
$^{1}$Institut d'Astrophysique et de G\'eophysique, Universit\'e de Li\`ege, 17, All\'ee du 6 Ao\^ut, B5c, B-4000 Sart Tilman, Belgium\\
$^{2}$Astronomical Institute 'Anton Pannekoek', University of Amsterdam, Postbus 94249, 1090 GE, Amsterdam, The Netherlands\\
$^{3}$UJF-Grenoble 1 / CNRS-INSU, Institut de Plan{\'e}tologie et d'Astrophysique de Grenoble (IPAG) UMR 5274, Grenoble, France\\
$^{4}$Royal Observatory of Belgium, Ringlaan 3, 1180, Brussels, Belgium}

\begin{document}

\date{Accepted . Received ; in original form }

\pagerange{\pageref{firstpage}--\pageref{lastpage}} \pubyear{2012}

\maketitle

\label{firstpage}

\begin{abstract}
Using optical long baseline interferometry, we resolved for the first time the two wide components of HD\,167971, a candidate hierarchical triple system known to efficiently accelerate particles. Our multi-epoch VLTI observations provide direct evidence for a gravitational link between the O8 supergiant and the close eclipsing O + O binary. The separation varies from 8 to 15\,mas over the three-year baseline of our observations, suggesting that the components evolve on a wide and very eccentric orbit (most probably $e>0.5$). These results provide evidence that the wide orbit revealed by our study is not coplanar with the orbit of the inner eclipsing binary. From our measurements of the near-infrared luminosity ratio, we constrain the spectral classification of the components in the close binary to be O6-O7, and confirm that these stars are likely main-sequence objects. Our results are discussed in the context of the bright non-thermal radio emission already reported for this system, and we provide arguments in favour of a maximum radio emission coincident with periastron passage. HD\,167971 turns out to be an efficient O-type particle accelerator that constitutes a valuable target for future high angular resolution radio imaging using VLBI facilities.
\end{abstract}


\section{Introduction}
Massive binaries made of O-type and/or Wolf-Rayet (WR) stars are known to be the scene of stellar wind collisions whose strength is intimately dependent on wind properties. Beside the hydrodynamic aspect of such a phenomenon \citep[e.g.][]{pit2009}, colliding-winds offer the opportunity to study higher level physical processes such as copious thermal X-ray emission \citep{SBP,PS,pitpar2010} or even particle acceleration mechanisms \citep{EU,BRwr,PD140art,debeckerreview}. The latter process considers a particle (electrons and hadrons) acceleration mechanism taking place in the presence of the strong shocks of the colliding-winds. Historically, the existence of a population of relativistic electrons has been revealed by the detection of non-thermal (synchrotron) radio emission coming from several massive stars. More recently a significant non-thermal high energy counterpart has been confirmed in the case of WR\,140 in hard X-rays \citep{sugawara140}, and even in $\gamma$-rays for $\eta$\,Car \citep{farnieretacar}.

\begin{table*} 
\begin{center}
\begin{minipage}{150mm}
\caption{Observing log for HD\,167971 and its calibrators.  \label{tab:log}}
\begin{tabular}{llllllll}
\hline
Epoch & Date & MJD & Star & Type & \# files & Instrument & Baseline \\
\hline
1 & May 17, 2008 & 54603.258 & HD\,167971 & SCI & 11 & AMBER & U1-U3-U4 \\
1 & May 17, 2008 & 54603.275 & HD\,179688 & CAL & 9 & AMBER & U1-U3-U4 \\
1 & May 17, 2008 & 54603.339 & HD\,167971 & SCI & 9 & AMBER & U1-U3-U4 \\
1 & May 17, 2008 & 54603.356 & HD\,179688 & CAL & 9 & AMBER & U1-U3-U4 \\
\hline
2 & July 16, 2008 & 54663.153 & HD\,167971 & SCI & 8 & AMBER & U1-U3-U4 \\
2 & July 16, 2008 & 54663.170 & HD\,179688 & CAL & 6 & AMBER & U1-U3-U4 \\
2 & July 17, 2008 & 54664.105 & HD\,154088 & CAL & 5 & AMBER & U1-U3-U4 \\
2 & July 17, 2008 & 54664.120 & HD\,167971 & SCI & 5 & AMBER & U1-U3-U4 \\
2 & July 17, 2008 & 54664.161 & HD\,179688 & CAL & 8 & AMBER & U1-U3-U4 \\
\hline
3 & March 29, 2011 & 55549.389 & HD\,167971 & SCI & 5 & AMBER & A0-G1-I1 \\
3 & March 29, 2011 & 55549.401 & HD\,166521 & CAL & 5 & AMBER & A0-G1-I1 \\
3 & March 29, 2011 & 55549.417 & HD\,167971 & SCI & 5 & AMBER & A0-G1-I1 \\
\hline
4 & August 9, 2011 & 55782.141 & HD\,166521 & CAL & 5 & PIONIER & A1-G1-I1-K0 \\
4 & August 9, 2011 & 55782.151 & HD\,167971 & SCI & 5 & PIONIER & A1-G1-I1-K0 \\
4 & August 9, 2011 & 55782.161 & HD\,166191 & CAL & 5 & PIONIER & A1-G1-I1-K0 \\
\hline
\end{tabular}
\end{minipage}
\end{center}
\end{table*}

So far, less than 40 massive stars revealed their capability to accelerate particles up to relativistic energies \citep{debeckerreview,benagliareview}. Even though it is now rather well-established that most of them are confirmed binaries, uncertainties remain concerning the multiplicity status of several targets. Among these, HD\,167971 deserves a particular attention. This system is known to be one of the brightest synchrotron radio emitters \citep{BAC,Blo167971}, clearly establishing its particle accelerator status. \citet{lei87} and \citet{DF1988} showed that it consists of a close O + O eclipsing binary (P\,=\,3.3213\,d) with a third more luminous O8I star. Considering the strong opacity of stellar winds to radio photons, the non-thermal radio emission is not expected to escape the wind-wind interaction region within the close binary \citep[see][for a discussion]{Blo167971}, suggesting electrons should be accelerated in a putative interaction region involving the third star. The study of Blomme et al.\ also revealed a radio light curve displaying long-term variations, compatible with a period of $\sim$\,20\,yr or more. HD\,167971 appears to be the brightest X-ray source in the NGC\,6604 open cluster, with a rather hard thermal emission spectrum compatible with pre-shock velocities too large to be attributed to the close binary colliding-winds \citep{DeB167971}. With a period of about 3.3\,d, the separation in the eclipsing binary is too short to allow stellar winds to reach their terminal velocities before collision, leading therefore to low post-shock temperatures unable to produce a significantly hard X-ray spectrum. The latter point suggests again an interaction between the winds of the close binary with that of the distant O8 supergiant, although its gravitational link with the close binary has yet to be demonstrated.

In this paper, we report on the results of long baseline interferometric observations of HD\,167971 performed with the Very Large Telescope Interferometer (VLTI) at ESO (European Southern Observatory, Chile). The observations and data processing are briefly described in Sect.\,\ref{obser}. The results are discussed in detail in Sect.\,\ref{discu} and we conclude in Sect.\,\ref{concl}.

\section{Observations and data processing}\label{obser}

\subsection{VLTI observations}
HD\,167971 was a target of the VLTI on May 17, 2008 and on July 16, 2008, using 3 Unit Telescopes (UT1-UT3-UT4) and the AMBER instrument \citep{amberart}. Our target was also observed on March 29, 2011 with AMBER and 3 Auxiliary Telescopes (ATs) in the A0-G1-K0 configuration, translating into baselines ranging between about 50 and 120\,m. The AMBER instrument was operated in the LR-HK mode, which provides a spectral resolution $R\simeq 35$ across the H and K bands. Calibrated visibility and closure phase values were computed using the latest public version of the AMBER data reduction package \citep[\texttt{amdlib} version 3.1,][]{Tatulli:2007,Chelli:2009} and the \texttt{yorick} interface provided by the Jean-Marie Mariotti Center\footnote{JMMC softwares: \texttt{http://www.jmmc.fr}}. The data were processed following the same procedure as described by \citet{hd93250vlti}. The accuracy of the absolute calibration was typically of 20\% for the visibilities and five degrees for the closure phases. 

On August 9, 2011, HD\,167971 was observed with the visitor-instrument PIONIER \citep{LeBouquin:2011} fed by all four ATs of VLTI in the A1-G1-I1-K0 configuration. PIONIER was operated with a low spectral resolution $R\approx15$, that is, three spectral channels across the H band. Calibrated visibility and closure phase values were computed using the latest version of the PIONIER data reduction package \citep[\texttt{pndrs} version 2.3,][]{LeBouquin:2011}. The accuracy was typically of 5\% for the visibilities and one degree for the closure phases. Table~\ref{tab:log} summarises all observations obtained for the four observing runs, including the observations of calibrator stars used to estimate the instrumental visibility and closure phase.

Special care was taken to retrieve the absolute sign of the closure phases for both instruments, which were previously calibrated with observations of known binaries.

\subsection{Binary modeling}

\begin{figure}
\begin{center}
\includegraphics[width=85mm]{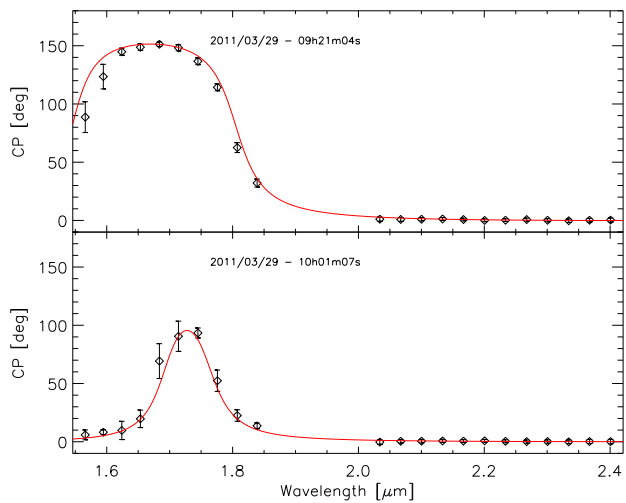}
\caption{Illustration of the closure phases obtained during the epoch 3 observing run on AMBER (diamonds with error bars), and of the best-fit binary model (solid line). The $\chi^2$ value associated to this fit is about 0.8. \label{fit}}
\end{center}
\end{figure}

The angular separation between the two stars in the close eclipsing binary is expected to be of the order of 0.1\,mas (assuming a circular orbit with a period of 3.3\,d), and is therefore seen as an unresolved point source by our interferometer. The O8I outer star is also seen as a point-like source, with an estimated angular diameter of 0.05\,mas based on the work of \citet{martins}. We therefore a priori expected to separate two point-like objects in our observations: the close binary and the O8 supergiant. For all four epochs, large closure phases and strong visibility variations with respect to wavelength, baselines and time are observed. This is illustrated in Fig.\,\ref{fit} for the closure phases at epoch 3. This suggests indeed a binary system with a flux ratio close to unity. We therefore decided to model the system by two unresolved stars with a constant flux ratio over the H and the K bands, a valid assumption for O+O binaries.

\begin{figure}
\begin{center}
\includegraphics[width=90mm]{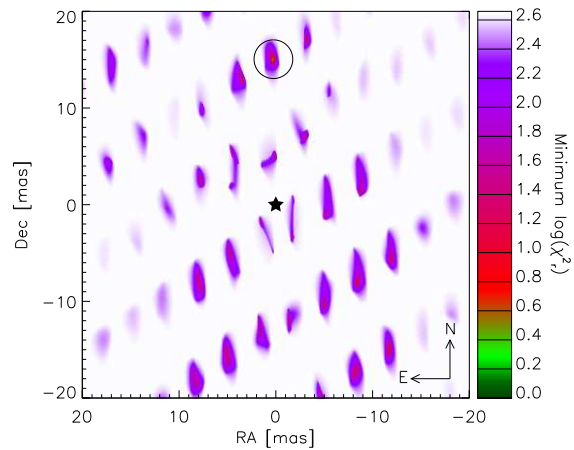}
\caption{Reduced $\chi^2$ map resulting from the fit of a binary model to the closure phases of the epoch 3 AMBER data set. The circle indicates the position of the sharply peaked minimum $\chi^2$ value. The figure appears in colour in the electronic version of the paper. \label{chi2}}
\end{center}
\end{figure}

We used the calibrated closure phases to derive the flux ratio and astrometry of the binary system, following the procedure presented in \citet{Absil:2011}. In short, we constructed binary models with various flux ratios and positions within the interferometric field-of-view ($\sim 40$\,mas on the UTs, $\sim 200$\,mas on the ATs at H band), and computed the reduced $\chi^2$ goodness of fit for all models to populate a $\chi^2$ cube. The global minimum, its significance and its uniqueness were then determined by inspecting the cube. For all epochs, a unique binary solution was found. The best fit to the data for epoch 3 and the resulting $\chi^2$ map are shown in Fig.\,\ref{fit} and Fig.\,\ref{chi2}, respectively. It is clear that the measurement of the binary separation is unambiguous: all locations are associated to $\chi^2 \ge 5$ except the one identified with a circle in Fig.\,\ref{chi2}, which has $\chi^2 \sim 0.8$.

The uncertainties on the best fit parameters were derived by inspecting the behavior of the reduced $\chi^2$ cube around the absolute minima \citep[see][for details]{Absil:2011}. The final error bars on the astrometry also take into account the estimated uncertainty on the spectral calibration of the two instruments (i.e., central wavelength of the spectral channels). It was conservatively set to $3.2\%$ in AMBER (twice the width of a spectral channel) and to $2\%$ in PIONIER \citep{LeBouquin:2011}. Finally, we checked that a similar analysis, performed on the calibrated squared visibilities instead of the closure phases, gives consistent results, although generally with larger error bars as discussed in \citet{Absil:2011}. The parameters of the best fit models for the four epochs are summarized in Table~\ref{results}.

\begin{table*} 
\begin{center}
\begin{minipage}{150mm}
\caption{Relative position of component B with respect to component A, and near-infrared luminosity ratios for the three observations (average over H and K bands for AMBER, and over the H band for PIONIER). The angular separation (S) and the position angle ($\theta$) are also given.  \label{results}}
\begin{tabular}{lllllll}
\hline
Epoch	& Date & $\delta$\,RA & $\delta$\,DEC & S & $\theta$ & L$_\mathrm{B,HK}$/L$_\mathrm{A,HK}$ \\
	&  & (mas) & (mas) & (mas) & (deg) &  \\
\hline
1 & May 17, 2008 & $8.28 \pm 0.27$ & $-2.53 \pm 0.09$ & $8.66 \pm 0.28$ & $107.0 \pm 1.1$ & $0.76 \pm 0.01$ \\
2 & July 16-17, 2008 & $8.78 \pm 0.28$ & $-1.03 \pm 0.05$ & $8.84 \pm 0.28$ & $96.7 \pm 0.6$ & $0.81 \pm 0.04$ \\
3 & March 29, 2011 & $0.28 \pm 0.03$ & $15.03 \pm 0.48$ & $15.03 \pm 0.48$ & $1.1 \pm 0.2$ & $0.83 \pm 0.01$ \\
4 & August 9, 2011 & $-1.53 \pm 0.03$ & $15.03 \pm 0.30$ & $15.11 \pm 0.30$ & $354.2 \pm 0.2$ & $0.75 \pm 0.01$ \\
\hline
\end{tabular}
\end{minipage}
\end{center}
\end{table*}

\section{Results and discussion}\label{discu}

\subsection{First evidence for a wide orbit}

\begin{figure}
\begin{center}
\includegraphics[width=85mm]{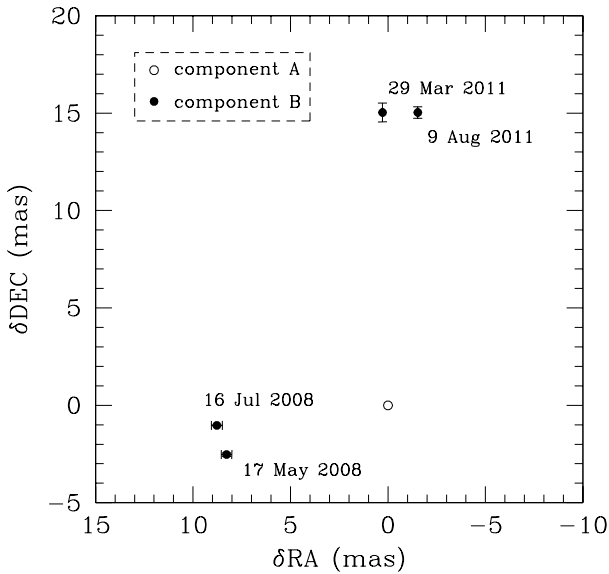}
\caption{Relative positions of components A and B in HD\,167971 at the four observation epochs. Component A, the brightest object, is arbitrarily located at coordinates (0,0).\label{visorbit}}
\end{center}
\end{figure}

The main result of our investigation is a separation of HD\,167971 into two components, respectively referred to as A and B. The brightest one, component A, should a priori be associated to the O8 supergiant (see Sect.\,\ref{comp} for a detailed discussion). The best-fit relative positions of the two resolved components in HD\,167971 are given in Table\,\ref{results}, and plotted in Fig.\,\ref{visorbit}. A second important result is the clear demonstration that the relative positions of components A and B changed significantly from one epoch to the other. In particular, it is noticeable that we detect a weak (but significant) change between observations 1 and 2, separated by only about 2 months. These facts provide for the first time compelling evidence that the third star is gravitationally bound to the close binary, definitely confirming the triple status of HD\,167971. Considering the lower limit on the orbital period suggested by radio light curves \citep[P\,$\sim$\,20\,yr,][]{Blo167971}, this strongly suggests a very large eccentricity for the orbit with a periastron passage rather close to the epochs 1 and 2.

The relation between periastron ($d_p$) and apastron ($d_a$) stellar separations, and eccentricity ($e$), is the following:
\begin{equation}
\frac{d_p}{d_a} = \frac{1 - e}{1 + e}
\end{equation} 
On the basis of the values of the relative separation at epochs 1 and 4, i.e., two extreme separations in our sampling, we derive an upper limit for the periastron over apastron separation ratio of about 0.6, translating into a lower limit on the eccentricity of about 0.25. It should be emphasized that this value is very conservative as the separation at epochs 3 and 4 strongly underestimate the expected separation at apastron. The configurations at four epochs illustrated in Fig.\,\ref{visorbit} suggest an actual separation at apastron at least twice as large as at epoch 4, converting into an eccentricity of at least 0.5. Even though our analysis led to valuable constraints on the properties of the wide orbit in HD\,167971, we note that the undetermined inclination, the uncomplete orbital coverage, and the lack of information from spectroscopic monitoring (especially close to periastron passage) inhibit our capability to derive the orbital parameters of the system. For this reason, we will refrain from speculating further on the orbital parameters of the wide orbit and will come back to this issue a soon as the adequate observational information (interferometric and spectroscopic) will be made available.

It is also interesting to note that our astrometric results for the wide orbit suggest a not so high inclination (defined as the angle between the line-of-sight and the normal to the orbital plane), although the eclipse measured for the close binary leads to an inclination angle for the short orbit of at least about 70$^\circ$ \citep{DF1988}. These results provide evidence for a significant inclination between the short and wide orbits in HD\,167971, i.e., the orbits are not coplanar. This fact is important for the issue of star formation scenarios depending on initial cluster conditions \citep[see e.g.][]{relori}.

\subsection{The components in HD\,167971}\label{comp}

Our interferometric measurements alone could not lift the ambiguity in the association of components A and B to the close binary and to the distant supergiant. However, \citet{lei87} reported that the O supergiant is the most luminous and dominates the optical and UV spectra. Even though our measurements were performed in the near-infrared domain, the rather similar effective temperatures of the three stars in the system strongly suggest that the luminosity ratio should go in the same direction in the infrared as in optical and UV domains. We will therefore consider that the supergiant is component A and that the close binary is component B.

\begin{table*}
\begin{center}
\begin{minipage}{150mm}
\caption{Potential stellar parameters for the stars in HD\,167971, assuming main-sequence ({\it upper part}) and giant ({\it lower part}) luminosity classes for the components in the close binary. We assumed the close binary to be made of two identical stars. \label{param}}
\begin{tabular}{llllllll}
\hline
Sp. type	& O5V & O5.5V & O6V & O6.5V & O7V & O7.5V & O8V \\
L$_\mathrm{BOL}$ (L$_\odot$)& 3.2\,$\times$\,10$^5$ & 2.6\,$\times$\,10$^5$ & 2.0\,$\times$\,10$^5$ & 1.6\,$\times$\,10$^5$ & 1.3\,$\times$\,10$^5$ & 1.0\,$\times$\,10$^5$ & 0.8\,$\times$\,10$^5$ \\
L$_\mathrm{B}$/L$_\mathrm{A}$ & 1.60 & 1.30 & 1.00 & 0.80 & 0.65 & 0.50 & 0.40 \\
\hline
Sp. type	& O5III & O5.5III & O6III & O6.5III & O7III & O7.5III & O8III \\
L$_\mathrm{BOL}$ (L$_\odot$)& 5.0\,$\times$\,10$^5$ & 4.3\,$\times$\,10$^5$ & 3.6\,$\times$\,10$^5$ & 3.1\,$\times$\,10$^5$ & 2.7\,$\times$\,10$^5$ & 2.3\,$\times$\,10$^5$ & 2.0\,$\times$\,10$^5$ \\
L$_\mathrm{B}$/L$_\mathrm{A}$ & 2.5 & 2.15 & 1.80 & 1.55 & 1.35 & 1.15 & 1.0 \\
\hline
\end{tabular}
\end{minipage}
\end{center}
\end{table*}

\begin{figure}
\begin{center}
\includegraphics[width=80mm]{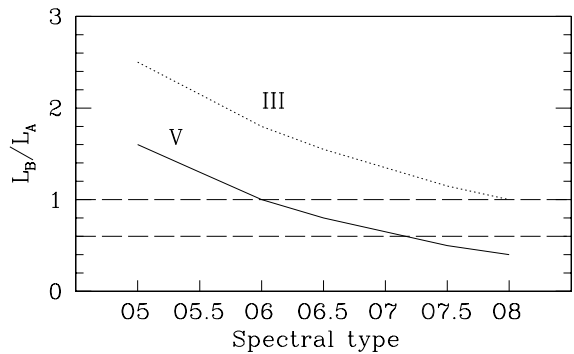}
\caption{Expected luminosity ratios assuming various spectral types, along with main-sequence (V) and giant (III) luminosity classes for the close binary components. The long-dashed horizontal lines encompass the ratio values compatible with our interferometric measurements, if one tolerates a 0.2 dex uncertainty.\label{lratiofig}}
\end{center}
\end{figure}

The luminosity ratio derived at every epoch can be discussed in order to derive additional constraints on the spectral classification of the stars in HD\,167971. Even though it is quite well-established that the third star should be an O8 supergiant, the spectral types of the components in the close binary are not fully determined. On the basis of the fundamental parameter values given by \citet{martins}, we derived rough estimates of the expected luminosity ratios assuming various spectral type combinations, and assuming the two components in the close binary to have the same spectral classification. According to the study of \citet{lei87}, the members of the close binary should be main-sequence stars, with spectral types ranging between O5 and O8. The expected bolometric luminosities and luminosity ratios, assuming an O8I third star (with L$_\mathrm{BOL}$\,=\,4\,$\times$\,10$^{5}$\,L$_\odot$, \citealt{martins}), are quoted in Table\,\ref{param}. The values given in Table\,\ref{results} for the measured luminosity ratio seem to significantly change from one epoch to the other. It should be emphasized that the eclipsing nature of the close binary (component B) may be responsible for a significant change of its brightness as a function of time, mostly considering the depth and duration of the eclipses reported for instance by \citet{DF1988}. However, the propagation of the uncertainties on the ephemeris of the eclipsing binary over more than 2000 cycles (between the photometric and interferometric observations) prevent us from determining the orbital phase of the close binary at the epoch of our interferometric observations. In addition, a slight difference may also appear between the three AMBER observations and the PIONIER measurement, as these instruments operate in different wavebands (HK and H, respectively). Assuming that luminosity ratios should be similar for both bolometric and K-band luminosities, our L$_\mathrm{B,HK}$/L$_\mathrm{A,HK} \sim 0.8$ measurements suggests spectral types for the stars in the close binary between O6 and O7 (typically, O6.5 with an uncertainty of half a spectral type) for main-sequence stars, as illustrated in Fig.\,\ref{lratiofig}\footnote{The uncertainty of 0.2\,dex taken into account in this figure could seem overestimated considering the error bars given in Table\,\ref{results} for the luminosity ratios, but we selected a comfortable confidence range in order to be sure to include the a priori unknown uncertainty on the theoretical luminosity ratio values, and therefore get rid of any overinterpretation of the results.}. For a giant luminosity class, the close binary components should not be earlier than O8, assuming that the third star is at least as luminous as the close binary. On the basis of these estimates we favor the main-sequence luminosity classes as suggested by \citet{lei87}, and we consider that the giant or even supergiant luminosity classes as claimed by \citet{DF1988} are unlikely. This result is also in agreement with the idea that the typical radius of evolved stars may be too large to fit within the small orbit of the eclipsing binary. Considering typical radii and masses for O6.5 stars given by \citet{martins}, and applying Kepler's law for a 3.3\,d circular orbit, we estimate that supergiants would be too large and that about 80\,$\%$ of the stellar separation would be occupied by the stars if they were giants. If main-sequence stars are assumed, we estimate that only about 50\,$\%$ of the stellar separation is filled by the stars. 
 
Assuming an (O6.5V + O6.5V) + O8I spectral classification for the stars in HD\,167971, we estimate the total visual absolute magnitude M$_V$\,=\,6.73 (for typical visual magnitudes as given by \citealt{martins}). For a V apparent magnitude of about 7.35 as reported by \citet{lei87} between the eclipses, and a color excess E(B -- V) of 0.9, we derive a distance modulus of 11.29, translating into a distance of about 1.8\,kpc. Considering the uncertainties, such a distance is in agreement with the distance of 1.7\,kpc reported by \citet{reipurth6604} for the NGC\,6604 open cluster, lending further support to the adopted spectral classification. 

We note that the additional information given here on the spectral classification of the components in HD\,167971, with respect to previously published results, should be considered with caution: stellar classification based on luminosities are indeed generally less accurate than classification approaches based on spectroscopic criteria. This issue would therefore strongly benefit from a long-term spectroscopic monitoring with application of state-of-the-art disentangling techniques to accurately classify individual components in the system, such as in the case of the hierarchical triple system HD\,150136 \citep{mahyhd150136}, provided the inclination of the wide orbit allows the measurement of significant radial velocity shifts of spectral lines of the three components.

\subsection{The radio emission in HD\,167971}\label{radio}

The radio light curve presented by \citet{Blo167971} shows a bright non-thermal radio emission whose strong variability is worth discussing in relation with the long period eccentric orbit. The demonstration of a gravitational connection between the O8 supergiant and the eclipsing binary provides a straightforwad interpretation scheme for the radio variability: the non-thermal radio photons should arise from an emission region coincident with the wind-wind interaction region in the wide binary, where relativistic electrons (and particles in general) are expected to be accelerated. The light curve presented by \citet{Blo167971} at 6\,cm is the one that benefits from the best time coverage. Its maximum occurred close to 1988-1989, and its minimum between 1995 and 1999. Assuming a period of about 20 years, the next maximum is expected to have occurred in 2008-2009, i.e., close to the time of our observations 1 and 2. In this context, it is tentative to attribute the maximum radio flux density to orbital phases close to periastron passage, i.e., when the separation is shorter. 

The non-thermal radio emission is indeed expected to be variable mainly for two reasons, respectively related to emission and absorption. First, the intrinsic synchrotron mechanism responsible for the non-thermal radio emission is more efficient when the stellar separation is shorter (typically, close to periastron passage). Synchrotron emission is indeed dependent on the local magnetic field strength ($B$), where relativistic electrons are accelerated: $ L_{\rm synch} \propto B^2 $. Considering the magnetic field strength dependence on the distance to the stars described for instance by \citet{UM}, $B$ is expected to decrease significantly as the stellar separation increases, leading to a maximum synchrotron emission at periastron. In addition, one should keep in mind that the kinetic power injected in the wind-wind interaction region, and consequently the power available for particle acceleration, is expected to decrease as the separation increases (see discussion in Sect.\,\ref{energy}). Second, radio emission from the colliding-winds region is also expected to vary significantly because of the phase-dependent free-free absorption due to the stellar winds material along the line of sight. In very eccentric orbits, the absorption is expected to be stronger close to periastron, when the synchrotron emission region is more likely to be located within the $\tau \sim 1$ radiosphere. The combination of these two effect is especially striking in the case of the long period Wolf-Rayet + O-type binary WR\,140 \citep{WhBe140}, where the radio maximum occurs slightly before periastron, just before the strong increase of the free-free absorption\footnote{We note that the situation may be somewhat more complex notably due to the Razin-Tsytovich effect, as illustrated in the case of the massive binary Cyg\,OB2\,\#8A \citep{Blocyg8a}, but the simple orbital phase dependence discussed here is correct to first order.}. In the case of HD\,167971, however, the impact of free-free absorption is a priori expected to be less spectacular: (i) the winds of the O-type stars in our target are not so efficient at absorbing radio photons as the WR wind in WR\,140; (ii) HD\,167971 is not so asymmetric as WR\,140 and the stagnation point in the wide orbit is more likely to be located close to mid-separation. As a result, we consider that the variability of the radio light curve in HD\,167971 should be dominated by the variation in the synchrotron emission, with a maximum rather close to periastron passage.

Finally, considering the rather wide orbit of the system whose wind-wind interaction is responsible for the non-thermal radio emission, along with the large brightness of the synchrotron source in the system, we claim that HD\,167971 would deserve high angular resolution radio imaging observations using Very Long Baseline Interferometer facilitites such as the VLBA, as already done in the case of a few systems such as WR\,140 \citep{Doug140}, WR\,146 \citep{oconnorwr146art}, WR\,147 \citep{williamswr147}, HD\,93129A \citep{benagliahd93129avlbi} and Cyg\,OB2\,\#5 \citep{ortizcyg5}. This would allow to spatially disentangle the thermal and non-thermal contributions in HD\,167971, and therefore more accurately quantify the synchrotron luminosity without contamination by the free-free (thermal) emission arising in individual stellar winds.

\subsection{Energy budget in HD\,167971}\label{energy}

The results detailed in previous sections allow us to give a few words on energetic considerations related to the physics of massive binaries, and in particular to particle acceleration and non-thermal processes in general. The available energy for physical processes related to the wind-wind interaction comes from the kinetic power of the colliding stellar winds. For a stellar wind with mass loss rate $\dot M$ and terminal velocity $V_\infty$, the kinetic power is $P_{kin} = 0.5\,{\dot M}\,V_\infty^2$. For a long period orbit as the one revealed by our interferometric measurements, the stellar winds are expected to have reached their terminal velocity before colliding. Considering the spectral types and luminosity classes discussed in Sect.\,\ref{comp}, and assuming terminal velocities of 3000\,km\,s$^{-1}$, we used the predicted mass loss rates given by \citet{mdotpred} based on stellar parameters from \citet{martins} to compute kinetic powers of about 1.9\,$\times$\,10$^{36}$ and 0.7\,$\times$\,10$^{36}$\,erg\,s$^{-1}$ respectively for components A and B. In order to evaluate the energy injected in the colliding-wind region, one has to consider the fractional solid angle of the wind-wind interaction region as seen from each star surface. Typically, one could consider that it amounts up to a few per cent. If a fraction of about 5\,$\%$ is assumed, the total energy per unit time injected in the wind-wind interaction by both stars is about 10$^{35}$\,erg\,s$^{-1}$. The efficiency of the conversion of collision mechanical energy to particle acceleration is expected to be of the order of one per cent, shared between electrons and hadrons (mostly protons), i.e., about 10$^{33}$\,erg\,s$^{-1}$.  The assumed value for the electron-to-hadron acceleration efficiency is still uncertain for colliding-wind binaries. Considering that this quantity is similar for environments accelerating cosmic-rays in general, it should be of the order of 1\,$\%$ \citep{Long,PD140art}. However, we should not reject the possibility that this efficiency ratio may be somewhat different, and possibly larger (1--10\,$\%$). Consequently, at the level of approximation of this discussion, we will consider that 10$^{31}$--10$^{32}$\,erg\,s$^{-1}$ is potentially available for non-thermal emission processes involving relativistic electrons (i.e., so-called leptonic processes).

In massive star environments, the most efficient non-thermal leptonic process is expected to be inverse Compton (IC) scattering (i.e., energy transfer from relativistic electrons to UV and visible photons that are up-scattered in the high energy domain). Considering the ratio between magnetic and radiation energy densities close to massive stars, IC scattering will be the dominant energy loss process for relativistic electrons \citep{PD140art,debeckerreview}. Typically, one could expect that at most one per cent of the energy transferred to relativistic electrons is converted into synchrotron radiation in the radio domain. As a result, the synchrotron radio luminosity should be of the order of 10$^{29}$--10$^{30}$\,erg\,s$^{-1}$. 

This number is worth comparing to observational quantities as reported in the study by \citet{Blo167971}. To do so, we need radio flux density measurements at the same epoch in order to derive a spectral index. We selected in the data published by Blomme et al. measurements obtained in February 1985\footnote{That date was selected as it corresponds to the epoch closest to the maximum radio emission for which reliable flux density measurements exist at two wavelengths in order to determine a sepctral index. At the maximum (1988-1989), flux density measurements exist only at 6\,cm.}, with measurements respectively of 7.4\,$\pm$\,0.4\,mJy at 2\,cm and 14.7\,$\pm$\,0.3\,mJy at 6\,cm. From these quantities, we determine a spectral index $\alpha \approx -0.6$, which is a comfortable non-thermal value\footnote{A thermal radio emission component from a spherically symmetric and homogeneous stellar wind is expected to be characterized by a spectral index close to 0.6, as demonstrated by \citet{WB} and \citet{PF}. A negative value for the spectral index is considered as a valuable tracer for non-thermal radio emission.}. We can therefore consider that the radio emission at the selected date is clearly dominated by synchrotron radiation, and that thermal contributions from the individual stellar winds contribute only a small fraction of the measured flux density at a given wavelength. Using equation (26) in \citet{debeckerreview} and assuming a distance of 1.7\,kpc for the NGC\,6604 open cluster harboring HD\,167971 \citep{reipurth6604}, we converted the flux density into an integrated radio luminosity of about 10$^{30}$\,erg\,s$^{-1}$. The latter value is close to the maximum expected on the basis of the above discussion. This suggests that the electron-to-hadron particle acceleration efficiency is rather high in HD\,167971, or that the energy injection in the particle acceleration mechanism is especially efficient. These considerations lend further support to the idea that HD\,167971 is an especially valuable target for the investigations of non-thermal processes in O-type multiple systems.
 
These estimates suggest finally that the remaining energy injected in the relativistic electron population (i.e., about 10$^{31}$--10$^{32}$\,erg\,s$^{-1}$) should be available for IC scattering, likely responsible for a weak hard X-ray emission (and potentially soft $\gamma$-rays). Considering the sensitivity of current high energy facilities, the detection of such an IC scattering emission is unlikely, and next generation hard X-ray facilities such as ASTRO-H (JAXA mission to be launched in 2014, \citealt{astroh}) are required to detect and quantify this emission process, and therefore derive valuable constraints on the non-thermal and electron-to-hadron efficiency.

\section{Conclusions}\label{concl}
Our analysis provided for the first time evidence that HD\,167971 is a gravitationally bound hierarchical triple system, with a very eccentric wide outer orbit. Using long baseline interferometry (with AMBER and PIONIER instruments at the VLTI), we resolved the system into two components: (i) the close eclipsing O-type binary and (ii) the additional O-type supergiant. The small fraction of the orbit covered by our observations suggests that the orbit between the two resolved components is wide, in agreement with the previously published lower limit on the orbital period of about 20\,yr, on the basis of the radio light curve. Our results suggest an eccentricity of at least 0.25 (and most probably $e > 0.5$), even though additional interferometric observations are required to ascertain its value.

Our measurements of the luminosity ratio allowed us to derive some constraints on the spectral classification of the members of the triple system. Considering that the brightest component is the O8I star identified by previous studies, our results favor an O6-O7 main-sequence classification for the two stars in the close binary (assuming these two stars are very similar, in agreement with previously published photometric results).

On the basis of these results, we can comfortably argue that the strong non-thermal radio emission associated to HD\,167971 should come from the wind-wind interaction region between the close binary and the additional supergiant. The rather wide orbit allows the bulk of the synchrotron radio emission to escape the stellar winds without undergoing strong free-free absorption. The large (though undetermined) eccentricity of the orbit provides a straightforward explanation for the strong variability of the non-thermal radio emission in HD\,167971. Assuming an orbital period of about 20\,yr, as suggested by previously published radio light curves, our results provide hints for a maximum radio emission occurring close to periastron passage. A more complete interferometric monitoring of HD\,167971 is however needed to establish that coincidence with more accuracy. We emphasize that our results lend additional support to the so called 'standard scenario' for non-thermal processes in massive star environments, that considers efficient particle acceleration in the colliding-winds region. Finally, we note that the wide orbit and the brightness of the synchrotron source in HD\,167971 designate it as a potential good target for future high resolution radio imaging observations of colliding-wind massive binaries.

\section*{Acknowledgments}

PIONIER is funded by the Universit\'e Joseph Fourier (UJF, Grenoble) through its Poles TUNES and SMING and the vice-president of research, the Institut de Plan\'etologie et d'Astrophysique de Grenoble, the ``Agence Nationale pour la Recherche'' with the program ANR EXOZODI, and the Institut National des Science de l'Univers (INSU) with the programs ``Programme National de Physique Stellaire'' and ``Programme National de Plan\'etologie''. The integrated optics beam combiner is the result of a collaboration between IPAG and CEA-LETI based on CNES R\&T funding. The authors want to warmly thank all the people involved in the VLTI project. MD thanks Dr. Gregor Rauw for a reading of a preliminary version of the manuscript, and Dr. Eric Gosset for fruitful and stimulating discussions on these results. Finally, it is a pleasure to thank the referee, Dr. Olivier Chesneau, for his positive and constructive comments that improved the manuscript. The SIMBAD database has been consulted fo
 r the bibliography. 

\bibliographystyle{mn2e}


\begin{thebibliography}{}

\bibitem[\protect\citeauthoryear{Absil et 
al.}{2011}]{Absil:2011} Absil O., et al., 2011, \aap, 535, A68

\bibitem[\protect\citeauthoryear{{Benaglia}}{{Benaglia}}{2010}]{benagliareview}
{Benaglia} P.,  2010, in {J.~Mart{\'{\i}}, P.~L.~Luque-Escamilla, \&
  J.~A.~Combi} ed., High Energy Phenomena in Massive Stars Vol.~422 of
  Astronomical Society of the Pacific Conference Series, {Non-Thermal Radio
  Emission from OB Stars: An Observer's View}.
p.~111

\bibitem[\protect\citeauthoryear{{Benaglia}, {Dougherty}, {Phillips},
  {Koribalski} \& {Tzioumis}}{{Benaglia} et~al.}{2010}]{benagliahd93129avlbi}
{Benaglia} P.,  {Dougherty} S.~M.,  {Phillips} C.,  {Koribalski} B.,
  {Tzioumis} T.,  2010, in Revista Mexicana de Astronomia y Astrofisica
  Conference Series Vol.~38, pp 41--43

\bibitem[\protect\citeauthoryear{{Benaglia} \& {Romero}}{{Benaglia} \&
  {Romero}}{2003}]{BRwr}
{Benaglia} P.,  {Romero} G.~E.,  2003, \aap, 399, 1121

\bibitem[\protect\citeauthoryear{{Bieging}, {Abbott} \& {Churchwell}}{{Bieging}
  et~al.}{1989}]{BAC}
{Bieging} J.~H.,  {Abbott} D.~C.,    {Churchwell} E.~B.,  1989, \apj, 340, 518

\bibitem[\protect\citeauthoryear{{Blomme}, {De Becker}, {Runacres}, {van Loo}
  \& {Setia Gunawan}}{{Blomme} et~al.}{2007}]{Blo167971}
{Blomme} R.,  {De Becker} M.,  {Runacres} M.~C.,  {van Loo} S.,    {Setia
  Gunawan} D.~Y.~A.,  2007, \aap, 464, 701
  
\bibitem[\protect\citeauthoryear{{Blomme}, {De Becker}, {Volpi}, \& {Rauw}}{{Blomme} et~al.}{2010}]{Blocyg8a}
{Blomme} R.,  {De Becker} M.,  {Volpi} D.,  {Rauw} G., 2010, \aap, 519, A111

\bibitem[\protect\citeauthoryear{Chelli, Utrera, 
\& Duvert}{2009}]{Chelli:2009} Chelli A., Utrera O.~H., Duvert G., 2009, A\&A, 502, 705 

\bibitem[\protect\citeauthoryear{{Davidge} \& {Forbes}}{{Davidge} \&
  {Forbes}}{1988}]{DF1988}
{Davidge} T.~J.,  {Forbes} D.,  1988, \mnras, 235, 797

\bibitem[\protect\citeauthoryear{{De Becker}}{{De
  Becker}}{2007}]{debeckerreview}
{De Becker} M.,  2007, \aapr, 14, 171

\bibitem[\protect\citeauthoryear{{De Becker}, {Rauw}, {Blomme}, {Pittard},
  {Stevens} \& {Runacres}}{{De Becker} et~al.}{2005}]{DeB167971}
{De Becker} M.,  {Rauw} G.,  {Blomme} R.,  {Pittard} J.~M.,  {Stevens} I.~R.,
   {Runacres} M.~C.,  2005, \aap, 437, 1029

\bibitem[\protect\citeauthoryear{{Dougherty}, {Beasley}, {Claussen}, {Zauderer}
  \& {Bolingbroke}}{{Dougherty} et~al.}{2005}]{Doug140}
{Dougherty} S.~M.,  {Beasley} A.~J.,  {Claussen} M.~J.,  {Zauderer} B.~A.,
  {Bolingbroke} N.~J.,  2005, \apj, 623, 447

\bibitem[\protect\citeauthoryear{{Eichler} \& {Usov}}{{Eichler} \&
  {Usov}}{1993}]{EU}
{Eichler} D.,  {Usov} V.,  1993, \apj, 402, 271

\bibitem[\protect\citeauthoryear{{Farnier}, {Walter} \& {Leyder}}{{Farnier}
  et~al.}{2011}]{farnieretacar}
{Farnier} C.,  {Walter} R.,    {Leyder} J.-C.,  2011, \aap, 526, A57

\bibitem[\protect\citeauthoryear{{Le Bouquin} {et~al.}(2011){Le Bouquin}, {Berger}, {Lazareff},
  {Zins}, {Haguenauer}, {Jocou}, {Kern}, {Millan-Gabet}, {Traub}, {Absil},
  {Augereau}, {Benisty}, {Blind}, {Bonfils}, {Bourget}, {Delboulbe},
  {Feautrier}, {Germain}, {Gitton}, {Gillier}, {Kiekebusch}, {Kluska},
  {Knudstrup}, {Labeye}, {Lizon}, {Monin}, {Magnard}, {Malbet}, {Maurel},
  {M{\'e}nard}, {Micallef}, {Michaud}, {Montagnier}, {Morel}, {Moulin},
  {Perraut}, {Popovic}, {Rabou}, {Rochat}, {Rojas}, {Roussel}, {Roux},
  {Stadler}, {Stefl}, {Tatulli}, \& {Ventura}}{{Le Bouquin} et~al.}{2011}]{LeBouquin:2011}
{Le Bouquin}, J.-B., {Berger}, J.-P., {Lazareff}, B., {et~al.} 2011, \aap, 535,
  A67

\bibitem[\protect\citeauthoryear{{Leitherer}, {Forbes}, {Gilmore}, {Hearnshaw},
  {Klare}, {Krautter}, {Mandel}, {Stahl}, {Strupat}, {Wolf}, {Zickgraf} \&
  {Zirbel}}{{Leitherer} et~al.}{1987}]{lei87}
{Leitherer} C.,  {Forbes} D.,  {Gilmore} A.~C.,  {Hearnshaw} J.,  {Klare} G.,
  {Krautter} J.,  {Mandel} H.,  {Stahl} O.,  {Strupat} W.,  {Wolf} B.,
  {Zickgraf} F.-J., {Zirbel} E.,  1987, \aap, 185, 121

\bibitem[\protect\citeauthoryear{{Longair}}{{Longair}}{1992}]{Long}
{Longair} M.~S.,  1992, {High energy astrophysics}.
Cambridge University Press, 2nd ed.

\bibitem[\protect\citeauthoryear{{Mahy}, {Gosset}, {Sana}, {Damerdji}, {De
  Becker}, {Rauw} \& {Nitschelm}}{{Mahy} et~al.}{2012}]{mahyhd150136}
{Mahy} L.,  {Gosset} E.,  {Sana} H.,  {Damerdji} Y.,  {De Becker} M.,  {Rauw}
  G.,    {Nitschelm} C.,  2012, \aap, in press

\bibitem[\protect\citeauthoryear{{Martins}, {Schaerer} \& {Hillier}}{{Martins}
  et~al.}{2005}]{martins}
{Martins} F.,  {Schaerer} D.,    {Hillier} D.~J.,  2005, \aap, 436, 1049

\bibitem[\protect\citeauthoryear{{Miyata}, {Tamura} \& {Kunieda}}{{Miyata}
  et~al.}{2011}]{astroh}
{Miyata} Y.,  {Tamura} K., {Kunieda} H.,  2011, in Society of Photo-Optical
  Instrumentation Engineers (SPIE) Conference Series Vol.~8147

\bibitem[\protect\citeauthoryear{{Muijres}, {Vink}, {de Koter}, {M{\"u}ller} \&
  {Langer}}{{Muijres} et~al.}{2012}]{mdotpred}
{Muijres} L.~E.,  {Vink} J.~S.,  {de Koter} A.,  {M{\"u}ller} P.~E.,
  {Langer} N.,  2012, \aap, 537, A37

\bibitem[\protect\citeauthoryear{{O'Connor}, {Dougherty}, {Pittard} \&
  {Williams}}{{O'Connor} et~al.}{2005}]{oconnorwr146art}
{O'Connor} E.,  {Dougherty} S.,  {Pittard} J.,    {Williams} P.,  2005, \jrasc,
  99, 142

\bibitem[\protect\citeauthoryear{{Ortiz-Le{\'o}n}, {Loinard},
  {Rodr{\'{\i}}guez}, {Mioduszewski} \& {Dzib}}{{Ortiz-Le{\'o}n}
  et~al.}{2011}]{ortizcyg5}
{Ortiz-Le{\'o}n} G.~N.,  {Loinard} L.,  {Rodr{\'{\i}}guez} L.~F.,
  {Mioduszewski} A.~J.,    {Dzib} S.~A.,  2011, \apj, 737, 30

\bibitem[\protect\citeauthoryear{{Panagia} \& {Felli}}{{Panagia} \&
  {Felli}}{1975}]{PF}
{Panagia} N.,  {Felli} M.,  1975, \aap, 39, 1

\bibitem[\protect\citeauthoryear{{Petrov}, {Malbet}, {Weigelt}, {Antonelli},
  {Beckmann}, {Bresson}, {Chelli}, {Dugu{\'e}}, {Duvert} \& {Gennari}}{{Petrov}
  et~al.}{2007}]{amberart}
{Petrov} R.~G.,  {Malbet} F.,  {Weigelt} G.,  {Antonelli} P.,  {Beckmann} U.,
  {Bresson} Y.,  {Chelli} A.,  et al.,
  2007, \aap, 464, 1

\bibitem[\protect\citeauthoryear{{Pittard}}{{Pittard}}{2009}]{pit2009}
{Pittard} J.~M.,  2009, \mnras, 396, 1743

\bibitem[\protect\citeauthoryear{{Pittard} \& {Dougherty}}{{Pittard} \&
  {Dougherty}}{2006}]{PD140art}
{Pittard} J.~M.,  {Dougherty} S.~M.,  2006, \mnras, 372, 801

\bibitem[\protect\citeauthoryear{Pittard \& Parkin}{Pittard \&
  Parkin}{2010}]{pitpar2010}
Pittard J.~M.,  Parkin E.~R.,  2010, MNRAS, 403, 1657

\bibitem[\protect\citeauthoryear{{Pittard} \& {Stevens}}{{Pittard} \&
  {Stevens}}{1997}]{PS}
{Pittard} J.~M.,  {Stevens} I.~R.,  1997, \mnras, 292, 298

\bibitem[\protect\citeauthoryear{{Reipurth}}{{Reipurth}}{2008}]{reipurth6604}
{Reipurth} B.,  2008, Handbook of Star Forming Regions, Vol.2, The Southern sky, p.~590

\bibitem[\protect\citeauthoryear{{Sana}, {Le Bouquin}, {De Becker}, {Berger},
  {de Koter} \& {M{\'e}rand}}{{Sana} et~al.}{2011}]{hd93250vlti}
{Sana} H.,  {Le Bouquin} J.-B.,  {De Becker} M.,  {Berger} J.-P.,  {de Koter}
  A.,    {M{\'e}rand} A.,  2011, \apjl, 740, L43+

\bibitem[\protect\citeauthoryear{{Stevens}, {Blondin} \& {Pollock}}{{Stevens}
  et~al.}{1992}]{SBP}
{Stevens} I.~R.,  {Blondin} J.~M.,    {Pollock} A.~M.~T.,  1992, \apj, 386, 265

\bibitem[\protect\citeauthoryear{{Sterzik} \& {Tokovinin}}{{Sterzik} \&
  {Tokovinin}}{2002}]{relori}
{Sterzik} M.~F.,  {Tokovinin} A.~A., 2002, \aap, 384, 1030

\bibitem[\protect\citeauthoryear{{Sugawara}, {Maeda}, {Tsuboi}, {Hamaguchi},
  {Corcoran}, {Pollock}, {Moffat}, {Williams}, {Dougherty} \&
  {Pittard}}{{Sugawara} et~al.}{2011}]{sugawara140}
{Sugawara} Y.,  {Maeda} Y.,  {Tsuboi} Y.,  {Hamaguchi} K.,  {Corcoran} M.,
  {Pollock} A.,  {Moffat} A.,  {Williams} P.,  {Dougherty} S.,    {Pittard} J.,
   2011, BSRSL, 80, 724

\bibitem[\protect\citeauthoryear{Tatulli et 
al.}{2007}]{Tatulli:2007} Tatulli E., et al., 2007, A\&A, 464, 29 

\bibitem[\protect\citeauthoryear{{Usov} \& {Melrose}}{{Usov} \&
  {Melrose}}{1992}]{UM}
{Usov} V.~V.,  {Melrose} D.~B.,  1992, \apj, 395, 575

\bibitem[\protect\citeauthoryear{{White} \& {Becker}}{{White} \&
  {Becker}}{1995}]{WhBe140}
{White} R.~L.,  {Becker} R.~H.,  1995, \apj, 451, 352

\bibitem[\protect\citeauthoryear{{Williams}, {Dougherty}, {Davis}, {van der
  Hucht}, {Bode} \& {Setia Gunawan}}{{Williams} et~al.}{1997}]{williamswr147}
{Williams} P.~M.,  {Dougherty} S.~M.,  {Davis} R.~J.,  {van der Hucht} K.~A.,
  {Bode} M.~F.,    {Setia Gunawan} D.~Y.~A.,  1997, \mnras, 289, 10

\bibitem[\protect\citeauthoryear{{Wright} \& {Barlow}}{{Wright} \&
  {Barlow}}{1975}]{WB}
{Wright} A.~E.,  {Barlow} M.~J.,  1975, \mnras, 170, 41

\end{thebibliography}

\bsp

\label{lastpage}

\end{document}